\documentclass[journal]{IEEEtran}

\IEEEoverridecommandlockouts
\pdfoutput=1
\usepackage[cmex10]{amsmath}
\usepackage{amssymb}
\usepackage{amsthm}
\usepackage{cases}
\usepackage{multirow}
\usepackage{empheq}

\usepackage{siunitx}

\usepackage[noadjust]{cite}
\usepackage{verbatim}

\usepackage{algorithm}
\usepackage{algpseudocode}

\ifCLASSINFOpdf
\usepackage[pdftex]{graphicx}
\else
\usepackage[dvips]{graphicx}
\fi
\usepackage[caption=false,font=footnotesize]{subfig}

\usepackage{color}
\usepackage{multirow}

\usepackage[hyphens]{url}

\newcommand{\vect}[1]{\mathbf{#1}}

\begin{document}
	
    \sloppy
	
    \title{Attention-aware Semantic Communications for Collaborative Inference}
	
    \author{
		\IEEEauthorblockN{Jiwoong Im, Nayoung Kwon, Taewoo Park, Jiheon Woo, Jaeho Lee, and Yongjune Kim}	\\		
    
    \thanks{

        
                
        J. Im, N. Kwon, T. Park, J. Woo, J. Lee, and Y. Kim are with the Department of Electrical Engineering, Pohang University of Science and Technology (POSTECH), Pohang 37673, South Korea (e-mail: \{jw3562, kwonna, parktaewoo, jhwoo1997, jaeho.lee, yongjune\}@postech.ac.kr). 
        J. Im and N. Kwon contributed equally. 
        
          
        }		
    }
	
	
    \maketitle
    	
    \begin{abstract}


    We propose a communication-efficient collaborative inference framework in the domain of edge inference, focusing on the efficient use of vision transformer (ViT) models.
    The partitioning strategy of conventional collaborative inference fails to reduce communication cost because of the inherent architecture of ViTs maintaining consistent layer dimensions across the entire transformer encoder. 
    Therefore, instead of employing the partitioning strategy, our framework utilizes a lightweight ViT model on the edge device, with the server deploying a complicated ViT model. 
    To enhance communication efficiency and achieve the classification accuracy of the server model, we propose two strategies: 1) attention-aware patch selection and 2) entropy-aware image transmission. 
    Attention-aware patch selection leverages the attention scores generated by the edge device's transformer encoder to identify and select the image patches critical for classification.
    This strategy enables the edge device to transmit only the essential patches to the server, significantly improving communication efficiency. 
    Entropy-aware image transmission uses min-entropy as a metric to accurately determine whether to depend on the lightweight model on the edge device or to request the inference from the server model. 
    In our framework, the lightweight ViT model on the edge device acts as a semantic encoder, efficiently identifying and selecting the crucial image information required for the classification task. 
    Our experiments demonstrate that the proposed collaborative inference framework can reduce communication overhead by \SI{68}{\%} with only a minimal loss in accuracy compared to the server model on the ImageNet dataset.       
    
    \end{abstract}  

	
    \section{Introduction}


    The rapid advancement of computational resources, coupled with the proliferation of massive datasets, has significantly enhanced the practicality of artificial intelligence (AI) services. 
    Integrating AI techniques with edge devices, including smartphones, wearable devices, and Internet of things (IoT) devices, seeks to seamlessly incorporate AI services into a wide range of daily life. 
    This effort to advance AI technologies in the domain of edge computing is commonly known as edge AI~\cite{Zhu2020toward,Li2020edge,Shao2020communication}. 


    An important research theme in edge AI is \emph{edge inference}, focused on efficiently executing inference tasks within the edge network~\cite{Li2020edge,Shao2020communication,Shlezinger2022collaborative,Lan2023progressive}. 
    Traditionally, raw data is sent from edge devices (clients) to a server, where a complicated model conducts the inference task, i.e., \emph{server-based inference}. 
    However, this method incurs significant communication overhead, particularly in scenarios dealing with large volumes of raw data~\cite{Shao2020communication,Shlezinger2022collaborative}. 
    An alternative is \emph{on-device inference}, which executes the inference task directly on resource-constrained devices, thereby minimizing communication costs. 
    However, this approach often leads to lower performance due to the limited computational capabilities of edge devices~\cite{Shao2020communication,Shlezinger2022collaborative}.    

    To address the dual challenges of excessive communication overhead and limited computational resources, the concept of \emph{collaborative inference} has been introduced~\cite{Li2020edge,Shao2020communication,Shlezinger2022collaborative,Lan2023progressive,Shao2022learning}. 
    This strategy involves dividing a deep neural network (DNN) model into separate parts for the edge device and the server. It leverages the model architecture of DNNs, where the dimensions of intermediate layers can be significantly smaller than the input dimensions.
    Within this framework, the edge device first uses its component to extract features from the raw data and then transmits them to the server. 
    As these extracted features are typically more compressed than the raw data, the communication cost can be aggressively reduced. 
    The server then utilizes these features and its portion of the model to determine the final inference result, which is sent back to the device~\cite{Lan2023progressive}. 
    The selection of the split point is critical as it significantly impacts the computational load on the edge device and the communication overhead~\cite{Shao2020communication}. 
    This approach is also known as \emph{split inference}~\cite{Huang2020dynamic,Lan2023progressive} and \emph{device-edge server co-inference}~\cite{Shao2020communication,Shao2022learning}. 
    Notably, collaborative inference is closely connected to \emph{semantic communications}~\cite{Zhang2022goal,Shi2021from,Lan2021what,Gunduz2022beyond,Xie2020lite,Kim2023distributed,Kim2020distributed}, considering that the extracted features are essentially semantic information tailored for the inference task.
    


    Transformers, originally developed for natural language processing (NLP)~\cite{Vaswani2017attention}, have been widely adopted across multiple domains. 
    Particularly, the vision transformer (ViT)~\cite{Dosovitskiy2021image,Touvron2021training} has demonstrated superior performance and efficiency in image classification tasks. 
    However, the deployment of ViTs on resource-constrained edge devices is challenging due to their substantial model size and intensive computational requirements~\cite{Mehta2022mobilevit}. 
        
    In collaborative inference scenarios, the strategy of partitioning ViT models fails to effectively reduce communication overhead. 
    This limitation stems from the inherent architecture of ViTs, which maintains consistent layer dimensions across the entire transformer encoder~\cite{Dosovitskiy2021image}, in contrast to DNN models whose intermediate layer dimensions can be significantly smaller than the raw data dimensions. 
    Hence, partitioning ViT models for collaborative inference cannot reduce communication overhead.
          

    In this paper, we propose a communication-efficient collaborative inference framework utilizing pre-trained ViT models. 
    Note that the collaborative inference in our work corresponds to the collaboration between an edge device and a server, rather than collaboration among multiple edge devices.
    Instead of partitioning a single model as in prior work~\cite{Li2020edge,Shao2020communication,Lan2023progressive,Shlezinger2022collaborative,Shao2022learning}, our approach involves the edge device operating a lightweight ViT model (e.g., DeiT-Tiny), while the server employs a more complex ViT model (e.g., DeiT-Base).
    As shown in Table~\ref{tab:model}, DeiT-Tiny (DeiT-Ti) is notably lightweight, making it suitable for edge deployment. 
    However, its classification accuracy is approximately \SI{10}{\%} lower than that of DeiT-Base (DeiT-B)~\cite{Touvron2021training}.

    \begin{table} [t]
    \renewcommand{\arraystretch}{1.2}
        \centering
        \caption{The Comparison of DeiT Model Complexity and Classification Accuracy on the ImageNet Dataset~\cite{Touvron2021training}}
        \begin{tabular}[width=0.2\textwidth]{|c|c|c|c|c|}
        \hline
            Model & Parameters  & Memory & 
            {FLOPs} & Classification\\ & (million) & {(MB)} & {(G)} & Accuracy (\%)\\\hline \hline
            DeiT-Tiny& 5 & 21.22 & 1.26 & 72.2\\
            DeiT-Small& 22 & 83.21 & 4.61 & 79.8\\
            DeiT-Base& 86 & 329.55 & 17.58 & 81.8\\ \hline
        \end{tabular}
        \label{tab:model}
    \end{table}

    %

    Our objective is to develop a collaborative inference strategy that achieves classification accuracy comparable to the server model while minimizing communication overhead between the edge device and the server. 
    This strategy is designed to leverage the strengths of both models: the efficiency and low resource demand of the tiny model on the edge device, and the higher classification accuracy of the base model on the server. 
    In our proposed framework, the edge device utilizes its tiny model to conduct an initial inference without transmitting the image to the server. 
    Subsequently, the edge device assesses whether to accept this initial inference or to send the image to the server for a more accurate inference using the base model. 
    By doing so, we aim to achieve an optimal trade-off between classification accuracy and communication cost in edge-server collaborative systems.

    To enhance communication efficiency in our framework, we propose two primary strategies: 1) attention-aware patch selection, which involves selectively transmitting only the most relevant patches of the image, and 2) entropy-aware image transmission, where the decision to transmit the image to the server is determined by the level of uncertainty or confidence in the edge device's initial inference. 
    \begin{itemize}
        \item \emph{Attention-aware patch selection}: In cases where the edge device needs to transmit the image to the server, our strategy is to transmit only the essential patches that are crucial for classification, rather than the entire image. 
        This selective transmission approach is guided by the \emph{attention scores} from the \emph{class token} to the image patches, as processed by the tiny model. 
        These attention scores indicate the relevance of each patch to the classification task. 
        We investigate several techniques to effectively select these important image patches using attention scores. 
        A crucial finding is that the tiny model is capable of accurately identifying the essential patches for the inference task, even when the client classifies the image incorrectly.
        Our experimental results validate that this approach enables the server model to maintain its classification accuracy, although it processes only selectively transmitted patches. 
        Hence, our attention-aware patch selection can effectively reduce the communication cost while achieving the classification accuracy of the server model. 
        
        \item \emph{Entropy-aware image transmission}: It is critical for the edge device to make an accurate decision between relying on its tiny model and requesting a more accurate inference from the base model on the server. 
        This decision significantly affects both the communication overhead and the classification accuracy. 
        Ideally, the edge device should transmit only those images incorrectly classified by the tiny model while avoiding the transmission of images where the initial inference is correct. 
        To facilitate this decision, we utilize the \emph{min-entropy} derived from the softmax output values of the classification head in the tiny ViT model. 
        Our experimental findings reveal that the decisions based on the min-entropy yield higher accuracy compared to those based on the Shannon entropy. 
        
    \end{itemize}


    In the context of semantic communications, our framework is aptly characterized as \emph{attention-aware semantic communications}. 
    The tiny model's \emph{transformer encoder} on the edge device acts as a \emph{semantic encoder}, particularly when the decision is made to transmit the image to the server. 
    The edge device leverages attention scores generated by the transformer encoder to identify the most essential image patches, which are critical for accurate classification. 
    Interestingly, our findings reveal that the tiny model functions effectively as a semantic encoder, in spite of its lower classification accuracy compared to the server model. 


    Moreover, the proposed collaborative inference framework offers the advantage of reducing the computational complexity on the server model, as the server's inference is conducted solely on the selected patches. 
    While our primary goal focuses on minimizing communication overhead between the edge device and the server, this framework also yields the ancillary benefit of server-side computational efficiency.     


    The rest of this paper is organized as follows. 
    Section~\ref{sec:related} offers a brief overview of the ViT and related work. 
    Section~\ref{sec:system} details our collaborative inference framework. 
    Section~\ref{sec:attention} and \ref{sec:entropy} present our main contributions, which include attention-aware patch selection and entropy-aware image transmission, respectively. 
    Section~\ref{sec:results} provides experimental results, followed by conclusions in Section~\ref{sec:conclusion}.

    \section{Backgrounds}\label{sec:related}

    \subsection{Vision Transformer}\label{sec:vit}     

\begin{figure}[t] 
        \centering
        \includegraphics[width=0.42\textwidth]{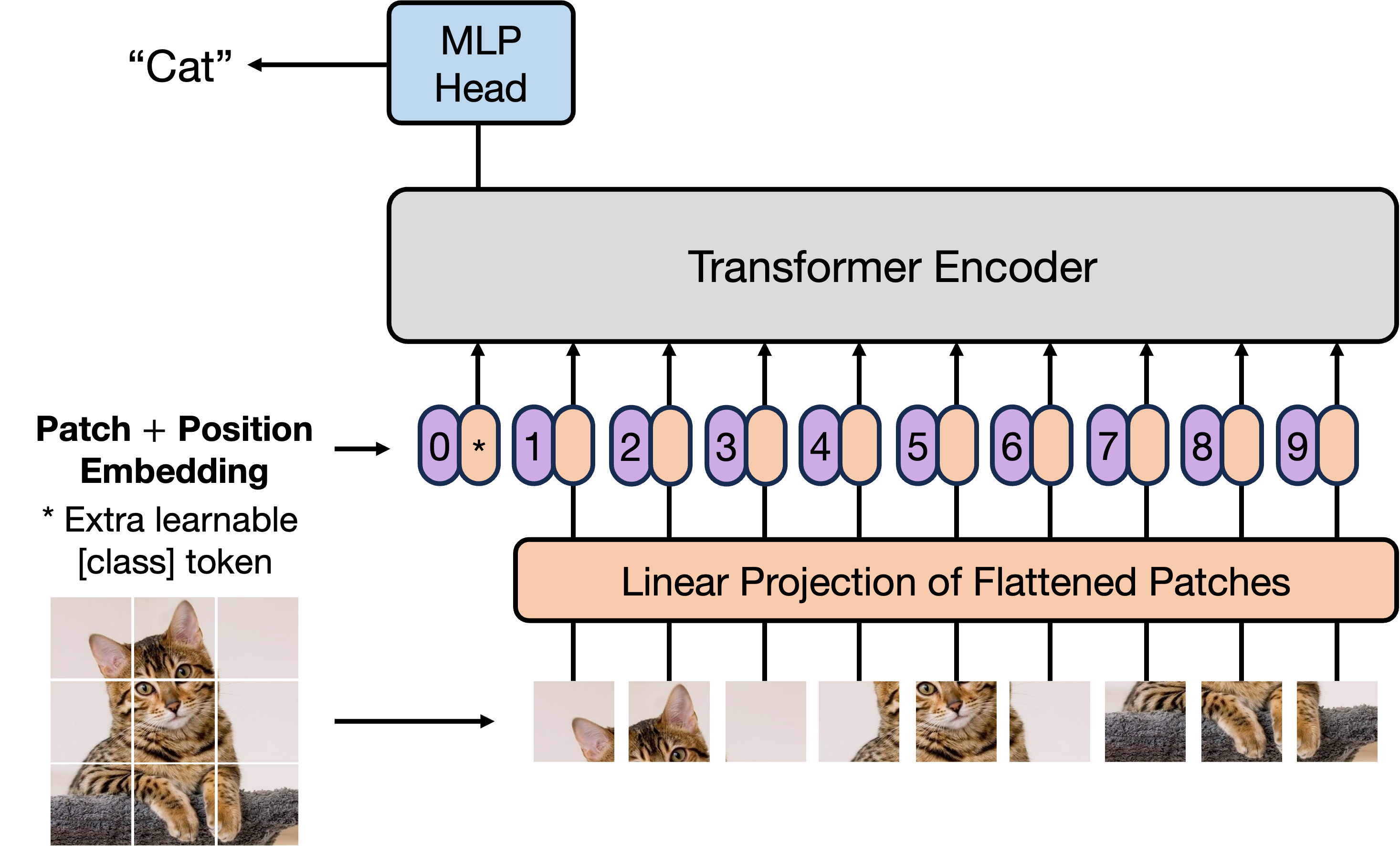}
        \caption{Overview of the ViT model~\cite{Dosovitskiy2021image}. 
        } 
        \label{fig:vit}
   \end{figure}

    The ViT~\cite{Dosovitskiy2021image} is a transformer-based model for computer vision tasks, setting a standard in vision models. 
    A simplified overview of the ViT model is shown in Fig.~\ref{fig:vit}. 
    An input image $\vect{x} \in \mathbb{R}^{H\times W \times C}$ is reshaped into a sequence of flattened 2D patches $\vect{x}_p \in \mathbb{R}^{N \times (P^2\cdot C)} $, where $(H, W)$, $C$, and $(P,P)$ denote the resolution of the original image, the number of channels, and the resolution of each image patch, respectively. 
    Note that $N = \frac{HW}{P^2}$ is the resulting number of patches. 
    These patches are then linearly projected to a consistent dimension $D$ across the transformer layers via $\vect{E} \in \mathbb{R}^{(P^2 \cdot C)\times D}$. 
    The input embedding of the ViT's transformer encoder $\vect{z}_0 \in \mathbb{R}^{(N+1) \times D}$ is given by
    \begin{equation}\label{eq:vit_input}
        \vect{z}_{0} = \left[\vect{x}_\text{cls};\vect{x}_p^{1} \mathbf{E}; \ldots ;\vect{x}_p^{N} \mathbf{E}\right] + \mathbf{E}_\text{pos},   
    \end{equation}
    where $\mathbf{E}_\text{pos}$ denotes the standard learnable position embedding. 
    The \emph{class token} $\vect{z}_0^0 = \vect{x}_\text{cls} \in \mathbb{R}^{1\times D}$ is particularly prepended to the sequence of embedded patches~\cite{Dosovitskiy2021image}. 
    This class token is crucial in classification tasks, serving as a key element in aggregating the information from the entire sequence of patches for the final classification output.
    
    The transformer encoder is composed of alternating layers of multi-head self-attention (MSA) and multi-layer perceptron (MLP) blocks as follows: 
    \begin{align}
        \vect{z}'_{l} &= \text{MSA}\left(\text{LN}\left(\vect{z}_{l-1}\right)\right) + \vect{z}_{l-1}, && l = 1,\ldots,L; \\
        \vect{z}_{l} &= \text{MLP}\left(\text{LN}\left(\vect{z}'_{l}\right)\right) + \vect{z}'_{l}, && l = 1,\ldots,L;  \\
        \vect{y} &= \text{LN}(\vect{z}_L^0),
    \end{align}
    where LN represents the layer normalization. 
    In particular, the image representation $\vect{y}$ is the encoder output of the class token $\vect{z}_0^0$. 
    This image representation $\vect{y}$ then serves as the input for the MLP head as shown in Fig.~\ref{fig:vit}. 
    

    In the MSA block for an input sequence $\vect{z} \in \mathbb{R}^{(N+1)\times D}$, a weighted sum of all values $\vect{v}$ is computed using query $\vect{q}$, key $\vect{k}$, and value $\vect{v}$, where $\vect{q}, \vect{k}, \vect{v} \in \mathbb{R}^{(N+1) \times D_h}$.
    The standard self-attention (SA) is formalized as follows~\cite{Vaswani2017attention,Dosovitskiy2021image}: 
    \begin{align}
        \left[\vect{q},\vect{k},\vect{v}\right] &= \vect{z}\mathbf{U}_{\vect{qkv}},   \\
        \mathbf{A} &= \text{softmax}\left(\frac{\vect{q}\vect{k}^\top}{\sqrt{D_h}}\right), \\
        \text{SA}(\vect{z}) &= \mathbf{A}\vect{v},
    \end{align}
    where $\mathbf{U}_{\vect{qkv}} \in \mathbb{R}^{D \times 3D_h}$ and $\mathbf{A} \in \mathbb{R}^{(N+1)\times(N+1)}$ denote the projection matrix and the attention weight matrix, respectively. 
    The MSA extends the standard self-attention (SA) by performing $H$ parallel SA operations (i.e., heads) and concatenating their outputs~\cite{Dosovitskiy2021image}:
    \begin{equation}
        \text{MSA}(\vect{z}) =\left[\text{SA}_1(\vect{z}),\cdots,\text{SA}_H(\vect{z})\right]\mathbf{U}_{\text{MSA}},
    \end{equation}
    where $\mathbf{U}_{\text{MSA}}$ is the projection matrix for the MSA output. 

    

    \subsection{Related Work}

        

    Prior work on collaborative inference has primarily focused on convolutional neural network (CNN) architectures~\cite{Li2020edge,Shao2020communication,Shlezinger2022collaborative,Jankowski2021wireless,Lan2023progressive,Shao2022learning}. 
    The exploration of collaborative inference with transformer models has been limited because of the consistent dimension of the MSA blocks in the transformer encoder. 

    Recent studies have delved into on-device inference utilizing lightweight ViT models~\cite{Mehta2022mobilevit,Pan2022edgevits,Li2022efficientformer}, yet achieving the accuracy of server models is challenging. 
    To improve the classification accuracy of the edge device, an ensemble strategy employing multiple edge devices was proposed in~\cite{Xu2023devit}. 
    In this approach, a central edge device disseminates raw data to nearby edge devices, which then conduct inference using compact ViT models and return their intermediate inference results to the central edge device. 
    The central edge device obtains the final inference outcome by aggregating the received intermediate results. However, this approach leads to increased communication overhead among the edge devices. 

    To reduce communication overhead, recent studies have introduced collaborative inference schemes utilizing ViT models on the server, namely masked autoencoder (MAE)-based offloading for transformer inference (MOT)~\cite{Liu2023efficient} and adaptive MOT (A-MOT)~\cite{Liu2024adaptive}. 
    In these approaches, the edge device selects image patches \emph{randomly} for transmission to the server. 
    The server then reconstructs the entire image using the decoder of MAE and performs classification on this reconstructed image.
    In spite of the benefit of reducing computational demands on edge devices, the classification accuracy is compromised by the random selection of image patches. 

    Transformers have been utilized in semantic communications for the transmission of text~\cite{Xie2021deep} and images~\cite{Yoo2023on}. 
    Nonetheless, these works do not pertain to classification tasks. 
    Our approach distinctively employs attention scores, particularly for the class token, setting our work apart from existing research in semantic communications.

    Recent studies have considered the importance of training data samples to enhance training performance in edge learning. 
    In~\cite{Liu2021data,Liu2021wireless}, the authors attempt to identify important training data samples and allocate more communication resources to these samples for improved communication efficiency. 
    The authors of \cite{Kook2024energy} select important features of training data samples to enhance communication efficiency during the training phase. 
    In contrast, we focus on communication-efficient collaborative inference, which specifically identifies important patches of test data samples and transmits only these essential patches. 
    Unlike these works that aim to enhance \emph{training performance}, our framework is designed to improve communication efficiency during the \emph{inference} phase.

    
 
    \section{Collaborative Inference Framework Based on Transformer Models}\label{sec:system} 

   \begin{figure*}[t] 
        \centering
        \includegraphics[width=0.7\textwidth]{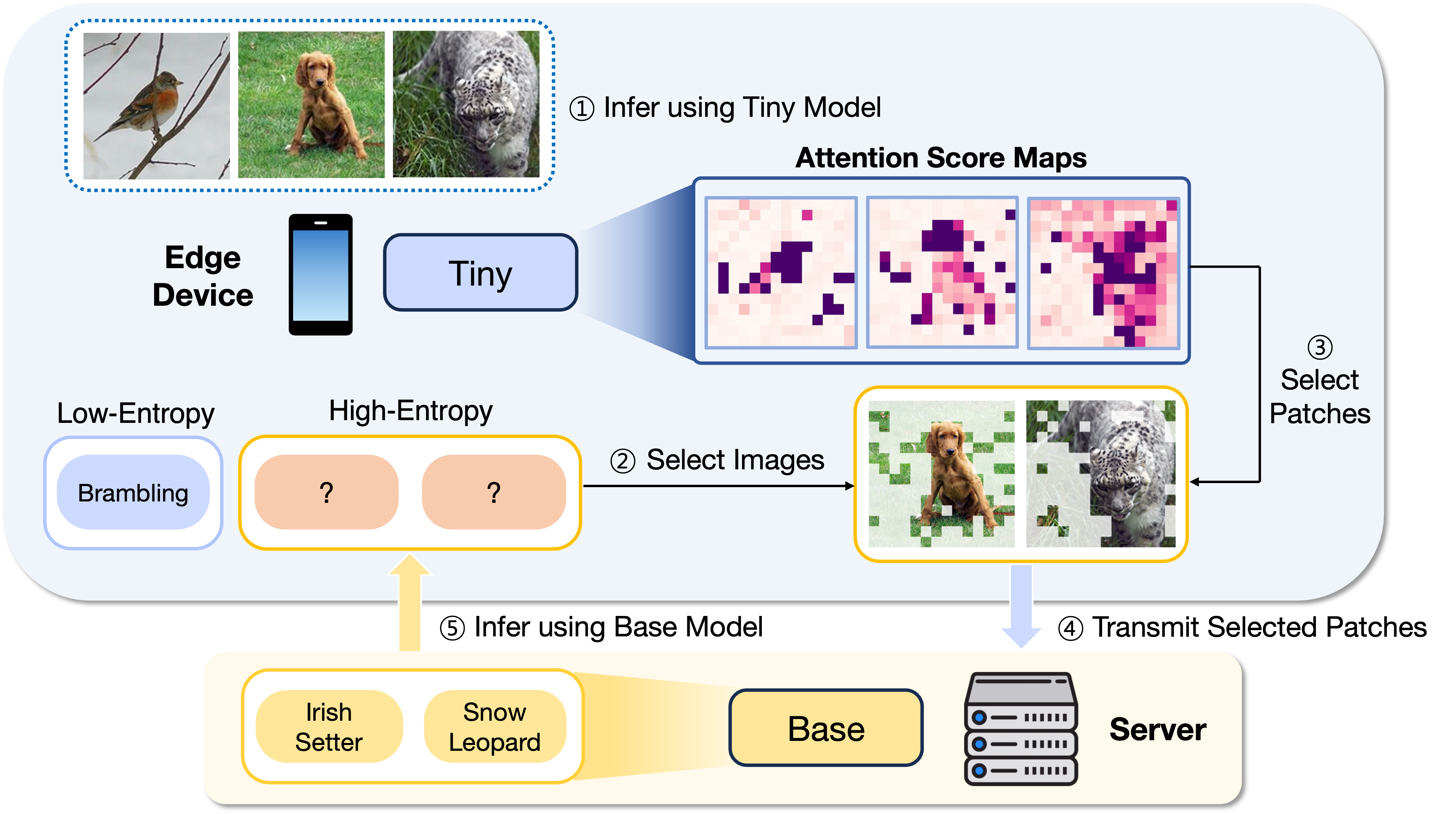}
        \caption{The proposed collaborative inference framework utilizing pre-trained ViT models: A lightweight model (e.g., DeiT-Tiny) on the edge device and a complicated model (DeiT-Base) on the server. The edge device (client) evaluates the uncertainty of its inference results by computing the entropy. If this entropy exceeds a predetermined threshold, the edge device selects the most important patches of the image based on the attention scores and transmits these to the server.} 
        \label{fig:framework}
        \vspace{-4mm}
   \end{figure*}

    We propose a collaborative inference framework that utilizes pre-trained ViT models. 
    This framework is designed to achieve server-level classification accuracy with minimized communication overhead between the edge device and the server.

    Due to the consistent layer dimensions of ViTs, 
    conventional methods of collaborative inference~\cite{Li2020edge,Shao2020communication,Shlezinger2022collaborative,Jankowski2021wireless,Lan2023progressive,Shao2022learning}, which typically partition a single DNN model, are ineffective at reducing communication costs for ViT models. 
    As a solution, we employ a lightweight ViT model (e.g., DeiT-Tiny) at the edge device, instead of splitting a complex ViT model (e.g., DeiT-Base), as depicted in Fig.~\ref{fig:framework}.  
    The proposed inference framework establishes an efficient collaborative protocol between the edge device and the server, aiming to achieve high classification accuracy of DeiT-Base model while significantly reducing communication overhead.

    In our collaborative inference framework, the edge device (client) first performs inference with its tiny model. The edge device then evaluates the entropy level of this initial inference. High entropy (or low confidence) necessitates transmitting the image to the server since it indicates that the tiny model's inference would be unreliable. In such instances, only essential patches for classification are transmitted instead of the entire image patches to minimize communication costs. The server, utilizing its complex ViT model, conducts inference based on these selected patches and sends its classification results back to the edge device, as shown in Fig.~\ref{fig:framework}. This process of selecting critical patches is governed by the proposed \emph{attention-aware patch selection} rule, elaborated in Section~\ref{sec:attention}. 
    
    If the initial inference's entropy is low, the edge device confirms its classification result without further interaction with the server, as shown in Fig.~\ref{fig:framework}. Reducing reliance on the server to reduce communication costs is achieved through \emph{entropy-aware image transmission} rule, detailed in Section~\ref{sec:entropy}. 
    By integrating these rules, our framework significantly lowers communication costs while maintaining classification accuracy comparable to the server model.  

   \begin{algorithm}[!t] 
    \caption{Proposed Collaborative Inference Framework} \label{algo:proposed}
    \textbf{Input:} Images $\{\vect{x}^{(1)}, \vect{x}^{(2)}, \ldots, \vect{x}^{(n)} \} $. \\
    \textbf{Output:} Classification results $\{ y^{(1)}, y^{(2)}, \ldots, y^{(n)} \} $.
    \begin{algorithmic}[1]
        \For {$i = 1:n $}
        \State $y^{(i)}_c \leftarrow f_c(\vect{x}^{(i)}) $ \Comment{Inference on edge device}
        \State Client computes entropy $g(\vect{x}^{(i)})$
            \If { $g(\vect{x}^{(i)}) \ge \eta$ }   
                \State $\widetilde{\vect{x}}^{(i)} \leftarrow \text{patch-selection}(\vect{x}^{(i)})$  
                \State{Client transmits $\widetilde{\vect{x}}^{(i)}$ to server }  
                \State ${y}^{(i)}_s \leftarrow f_s(\widetilde{\vect{x}}^{(i)}) $ \Comment{Inference on server}
                \State {Server transmits ${y}^{(i)}_s$ to client}
                \State {${y}^{(i)}_c \leftarrow{y}^{(i)}_s$}
            \EndIf
        \EndFor
    \end{algorithmic}
    \end{algorithm}

    The steps of the proposed collaborative inference are outlined in  Algorithm~\ref{algo:proposed}. Here, Step 2 and Step 3 involve computing the initial inference result $f_c(\vect{x}^{(i)})$ and its entropy $g(\vect{x}^{(i)})$, respectively.  
    If the entropy is below a given threshold $\eta$, then $f_c(\vect{x}^{(i)})$ is deemed the final classification outcome. 
    In cases of higher entropy, as identified in Step 4, the client selects and transmits only essential patches to the server at Step 5 and 6, effectively lowering communication costs by ensuring $\dim(\widetilde{\vect{x}}^{(i)}) < \dim(\vect{x}^{(i)})$. 
    At Step 7, the server conducts inference on these selected patches, producing the result $f_s(\widetilde{\vect{x}}^{(i)})$, which is then sent back to the client at Step 8. 

    The proposed collaborative inference framework can reduce the computational complexity for the server model by limiting the inference process to only the selected patches. 
    The computational complexity of DeiT-Base, according to~\cite{Chang2023making,Son2023maskedkd}, is given by
    \begin{equation} \label{eq:complexity}
        144 N D^2 + 24 N^2 D, 
    \end{equation}
    where $N$ is the number of patches. 
    Assuming the number of patches of $\widetilde{\vect{x}}^{(i)}$ is represented by $\widetilde{N}$ such that $\widetilde{N} < N$, our framework not only reduces communication overhead but also enhances computational efficiency on the server side. 
    This results in a significant secondary benefit of our collaborative inference framework. 
    
    \section{Attention-aware Patch Selection}\label{sec:attention} 

    %

    \begin{figure}[t] 
      \centering
      \begin{tabular}{cc}
        \includegraphics[width=0.3\linewidth]{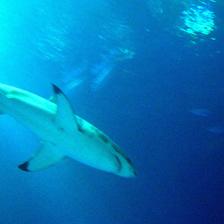} & 
        \includegraphics[width=0.3\linewidth]{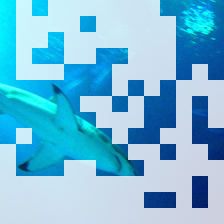}        
     \end{tabular}
      \caption{The visualization of attention-aware patch selection. The left is an image of the ImageNet dataset that the client model (DeiT-Tiny) inaccurately classifies as `Hammerhead Shark'. The right shows the selected patches by \emph{attention-aware patch selection}. These selected patches allow the server model (DeiT-Base) to correctly classify the image as `White Shark'.}
      \label{fig:motivation}
   \end{figure}     

       \begin{figure}[t] 
        \centering
        \begin{tabular}{ccc}
        \includegraphics[width=0.28\linewidth]{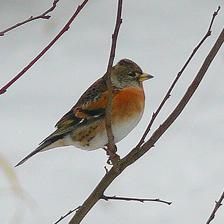} & 
        \includegraphics[width=0.28\linewidth]{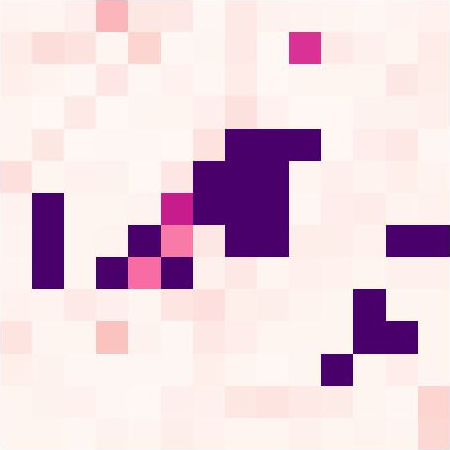} &
        \includegraphics[width=0.28\linewidth]{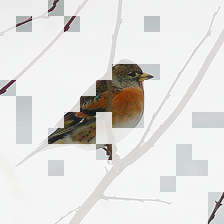}\\
        \includegraphics[width=0.28\linewidth]{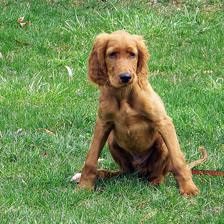} & 
        \includegraphics[width=0.28\linewidth]{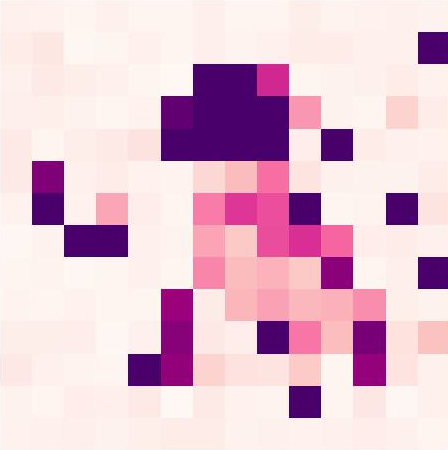} &
        \includegraphics[width=0.28\linewidth]{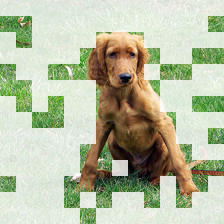}\\
        \includegraphics[width=0.28\linewidth]{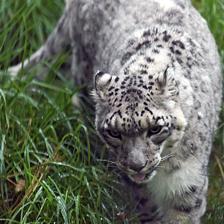} &
        \includegraphics[width=0.28\linewidth]{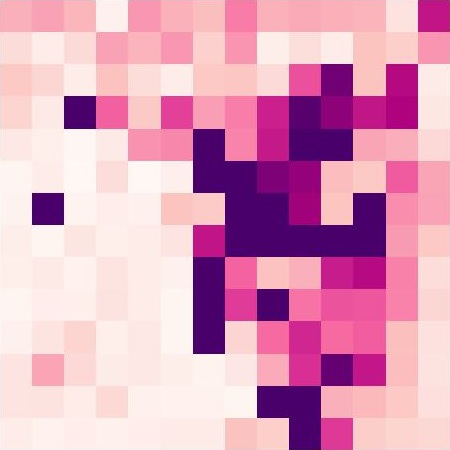} &
        \includegraphics[width=0.28\linewidth]{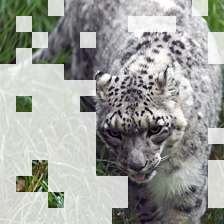} \\     
    \end{tabular}       
    \caption{The visualization of images and attention scores. In the left column, we have the original images of the ImageNet dataset. The middle column displays the attention score maps generated by DeiT-Tiny. The right column shows the patches selected by the attention-sum threshold selection. The images are labeled as `Brambling', `Irish Setter', and `Snow Leopard', respectively.}
    \label{fig:attention}
    \vspace{-4mm}
    \end{figure}

    This section introduces our attention-aware patch selection method, motivated by an intriguing observation: the tiny ViT model is capable of identifying the essential patches for classification, even when its classification is incorrect (see Fig.~\ref{fig:motivation}). 
    Consequently, the tiny model on the edge device acts as a \emph{semantic encoder}, effectively extracting essential information for the classification task. 
    
    To enhance communication efficiency and classification accuracy, we address two key questions: 1) how to accurately quantify the importance of each patch and 2) how to determine the optimal number of selected patches. 


    \subsection{Quantifying Patch Importance} \label{sec:quantifying}

    \begin{figure}[t] 
        \centering     
        \begin{tabular}{ccc}
        \includegraphics[width=0.28\linewidth]{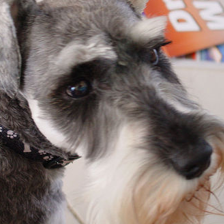} & 
        \includegraphics[width=0.28\linewidth]{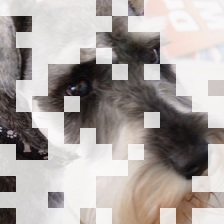} & 
        \includegraphics[width=0.28\linewidth]{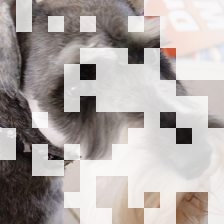} \\
        \includegraphics[width=0.28\linewidth]{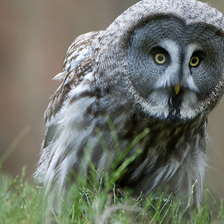} & 
        \includegraphics[width=0.28\linewidth]{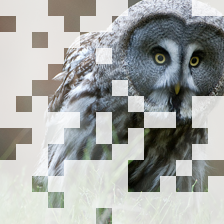} & 
        \includegraphics[width=0.28\linewidth]{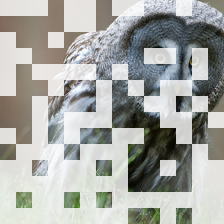} \\
        \includegraphics[width=0.28\linewidth]{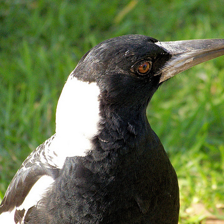} & 
        \includegraphics[width=0.28\linewidth]{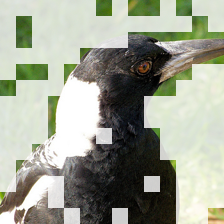} & 
        \includegraphics[width=0.28\linewidth]{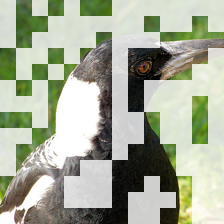} \\
        \end{tabular}       
    \caption{The comparison of image patches selected by DeiT-Tiny (middle column) and DeiT-Base (right column). The same number of patches are selected according to the mean attention scores. The left column displays the original images of the ImageNet dataset. The images are labeled as `Miniature Schnauzer', `Great Gray Owl', and `Magpie', respectively.}
    \label{fig:attention_model}
    \end{figure}

    To quantify the importance of each patch for classification, we utilize the \emph{attention scores} generated by the SA mechanism. 
    The attention score for the class token in a single-head attention is calculated as follows: 
    \begin{equation} \label{eq:attention}
        \vect{a} = \text{softmax}\left(\frac{\vect{q}_\text{cls} \vect{k}_p^\top}{\sqrt{D_h}}\right),
    \end{equation}
    where $\vect{q}_\text{cls} \in \mathbb{R}^{1 \times D_h}$ represents the query for the class token of the last layer and  $\vect{k}_p \in \mathbb{R}^{N \times D_h}$ denotes the keys corresponding to the image patches in the last layer.  
    The \emph{mean attention score} is then obtained by averaging the attention scores from all multi-heads.   

    Our experimental findings indicate that the mean attention scores, as computed by the tiny model, effectively assess the significance of each patch in contributing to the classification task. 
    Fig.~\ref{fig:attention} presents a side-by-side comparison of ImageNet dataset images (left column) and their corresponding attention score maps (middle column). 
    These maps clearly reveal that patches crucial for classification are distinguished by higher attention scores, setting them apart from less critical areas, such as background patches, which receive lower attention scores.


    This observation supports that the tiny model on the edge device is adept at identifying and selecting the most informative patches for classification.
    Within ViT models, the class token aggregates information from other tokens (image patches) via the attention mechanism. 
    For the final classification, the ViT relies on the MLP head, which considers only the information associated with the class token from the last layer, disregarding any other inputs.
    Thus, the attention score as defined in \eqref{eq:attention} serves as a key metric for quantifying the contributions of individual image patches to the class token. 
    In the context of semantic communications, the tiny model of the edge device acts as a \emph{semantic encoder}, tasked with extracting essential information for the classification task. 
    This role aligns with the broader objectives of semantic communications to emphasize meaning and relevance in the transmission of information~\cite{Gunduz2022beyond,Shi2021from,Lan2021what}

\begin{figure}[t] 
        \centering     
        \begin{tabular}{ccc}
        \includegraphics[width=0.28\linewidth]{fig/visual/9818.png} & 
        \includegraphics[width=0.28\linewidth]{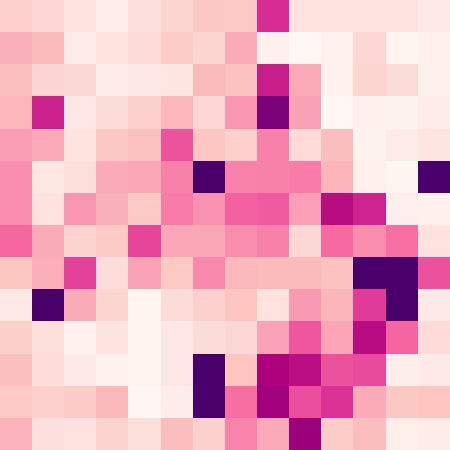} & 
        \includegraphics[width=0.28\linewidth]{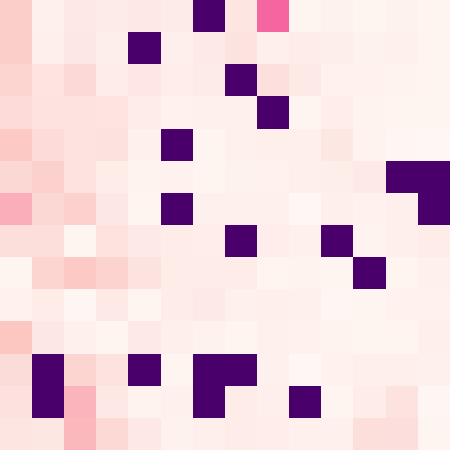} \\
        \includegraphics[width=0.28\linewidth]{fig/visual/1200.png} & 
        \includegraphics[width=0.28\linewidth]{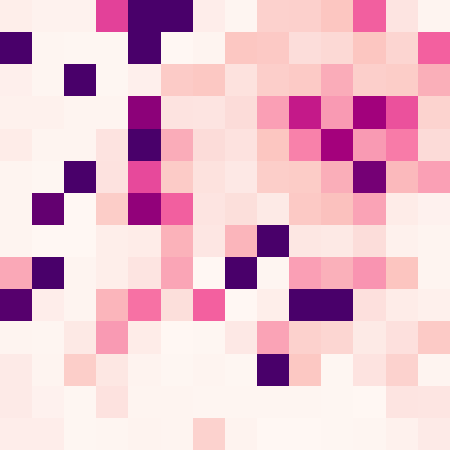} & 
        \includegraphics[width=0.28\linewidth]{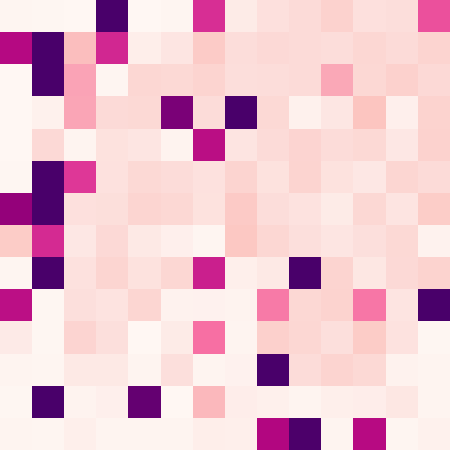} \\
        \includegraphics[width=0.28\linewidth]{fig/visual/928.png} & 
        \includegraphics[width=0.28\linewidth]{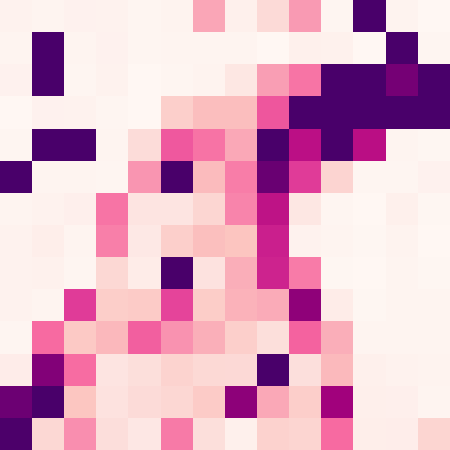} & 
        \includegraphics[width=0.28\linewidth]{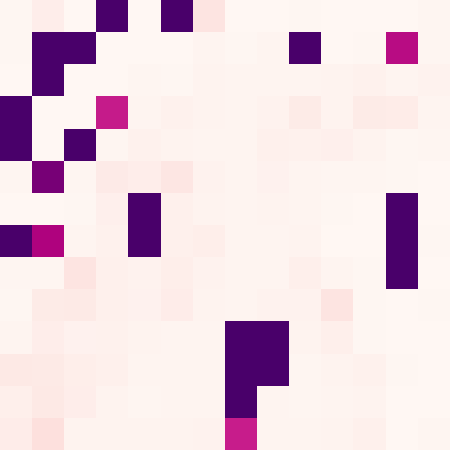} \\
        \end{tabular}
    \caption{The comparison of attention score maps computed by DeiT-Tiny (middle column) and DeiT-Base (right column). The left column displays the original images of the ImageNet dataset. The images are labeled as `Miniature Schnauzer', `Great Gray Owl', and `Magpie', respectively.}
    \label{fig:attention_map_model}
    \vspace{-4mm}
    \end{figure}

    An interesting finding is that DeiT-Tiny can act more effectively as a semantic encoder than DeiT-Base in spite of its inferior classification accuracy. 
    Fig.~\ref{fig:attention_model} compares the patches selected by DeiT-Tiny (middle column) and those by DeiT-Base (right column), showing DeiT-Tiny's superior ability to discard irrelevant image patches.
    This seemingly contradictory finding can be elucidated by the insights from recent work~\cite{Darcet2024vision}, which shows that large ViT models tend to allocate high attention scores to less informative background areas.
    It is because the large models adeptly identify patches with minimal information, such as background areas, repurposing the corresponding patches to assimilate global image information while neglecting their spatial information. 
    High attention scores are allocated to these repurposed patches containing global information, particularly in the background areas, as shown in Fig.~\ref{fig:attention_map_model}.
    Although this strategy enhances the classification accuracy of larger ViT models, it compromises their effectiveness as semantic encoders. 
    The experimental results on how model complexity affects the efficacy of semantic encoders are presented in Section~\ref{sec:semantic_enc}.
    
    The mean attention score from the last layer has previously been leveraged for purposes such as model interpretation~\cite{Caron2021emerging,Chefer2022optimizing} and knowledge distillation~\cite{Son2023maskedkd}.
    Our approach distinctively utilizes the mean attention score to reduce communication costs within our collaborative inference framework, differentiating our methodology from previous applications.
    Additionally, while \emph{attention rollout}~\cite{Abnar2020quantifying} is an established technique for interpreting transformer models, we opt for the mean attention score. 
    This decision is based on the observation that attention rollout tends to produce more uniform attention scores, which do not align well with our objective of attention-aware patch selection. 
    The experimental evidence supporting this decision is provided in Section~\ref{sec:patch_selection}.




    \subsection{Patch Selection Rule}


    In this subsection, we investigate the patch selection rule utilizing mean attention scores. 
    Selecting an appropriate number of patches for transmission to the server is crucial, as this directly impacts both communication costs and classification accuracy. 
    Our goal is to transmit the fewest possible patches to the server to minimize communication overhead. 
    However, this approach poses a trade-off, as reducing the number of transmitted patches can limit the information available to the server model, potentially lowering classification accuracy.
    
\begin{figure}[t] 
        \centering        
        \includegraphics[width=0.4\textwidth]{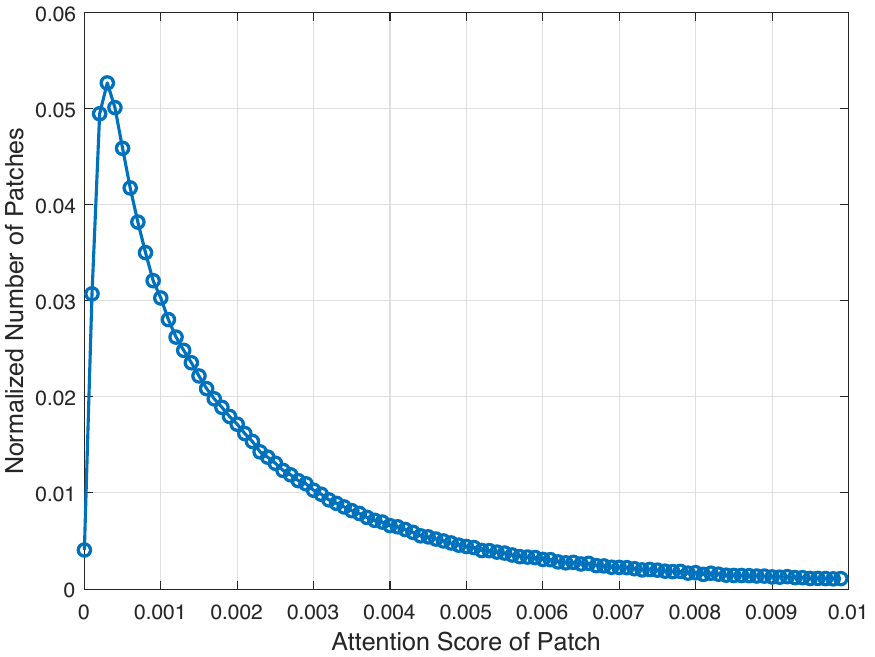}
        \caption{The normalized histogram of the mean attention scores obtained by DeiT-Tiny on the ImageNet dataset.}
        \label{fig:score_hist}
        \vspace{-4mm}
    \end{figure}

    The distribution of attention scores reveals that most patches are assigned low values, as shown in Fig.~\ref{fig:score_hist}.
    This property enables a significant reduction in the number of transmitted patches without affecting classification accuracy.
    By taking advantage of this beneficial property, we explore the following patch selection rules: 
    \begin{itemize}
        \item \emph{Top-$k$ selection}: Selecting the top-$k$ patches that have the highest attention scores.
        \item \emph{Attention threshold selection}: Selecting the patches whose attention scores exceed a predefined threshold $\delta$.
        \item \emph{Attention-sum threshold selection}: Selecting the patches with the highest attention scores until their cumulative attention sum reaches a predetermined threshold $\delta_{\text{sum}}$. 
    \end{itemize}


    The top-$k$ selection method selects a fixed number of patches based on the highest attention scores, resulting in consistent communication costs for all input images. 
    Nonetheless, it overlooks the variation in attention score distribution among different images. 
    Fig.~\ref{fig:attention} shows that the quantity of patches essential for classification can vary across images, with important information typically concentrated on the class object, where the highest attention scores are located. 
    Consequently, to achieve optimal classification accuracy, the number of selected patches should be tailored to the size of the object within each image. 

    Both the attention threshold selection and attention-sum threshold selection methods provide the capability to adjust the number of selected patches for transmission, making them more adaptable than the top-$k$ selection method. 
    For instance, as shown in Fig.~\ref{fig:attention}, the numbers of selected patches for `Brambling' and `Snow Leopard' are $35$ and $124$, respectively, by using the attention-sum threshold selection with $\delta_{\text{sum}} = 0.94$. 
    Adopting this method ensures the transmission of a consistent sum of attention scores, effectively lowering the risk of omitting crucial information.
    The experimental results show that the attention threshold selection and attention-sum threshold selection methods outperform the top-$k$ selection method in achieving an optimal trade-off between classification accuracy and communication efficiency, as detailed in Section~\ref{sec:patch_selection}.

    \section{Entropy-aware Image Transmission} \label{sec:entropy}

    This section delves into entropy-aware image transmission, a strategy aimed at reducing communication overhead by considering the varied classification difficulty inherent to different images.
    For less complex images, the edge device's initial inference may be accurate enough, eliminating the need for further interaction with the server. 
    In contrast, more intricate images necessitate more accurate classification from the server model, leading to increased communication overhead.
    It is critical for the edge device to make an accurate decision between relying on its initial inference and requesting more accurate classification from the server model.

    Even though the edge device cannot ascertain the correctness of its initial inference, it can estimate the inference's confidence through the softmax output values of the MLP classification head. 
    This softmax output can be interpreted as the posterior probability $p_\theta(y|\vect{x})$, where $y$ denotes the class label and $\theta$ denotes the tiny model. 
    Then, we set an entropy function $g: \mathbb{R} ^L \rightarrow \mathbb{R}$, where $L$ denotes the number of class labels. The client requests more accurate inference results from the server if:
    \begin{equation} \label{eq:entropy}
        g(\vect{x}) \ge \eta,
    \end{equation}
    where $\eta$ denotes a predetermined threshold.
    
    To assess the confidence of the client's inference, we consider two exemplary entropy measures: 1) Shannon entropy and 2) min-entropy, with their respective thresholds. 
        
    The Shannon entropy, a widely used metric for quantifying uncertainty \cite{Shannon1948mathematical}, is calculated by
    \begin{equation}
        g_s(\vect{x}) = -\sum_{y\in\mathcal{Y}} p_\theta(y\vert\vect{x})\log_2{p_\theta( y\vert\vect{x})},
    \end{equation}
    where $\mathcal{Y}$ denotes the set of all possible class labels. 
    High Shannon entropy indicates that the given image $\vect{x}$ is challenging for the tiny model to classify accurately. 
    Therefore, if $g_s(\vect{x}) \ge \eta_s$, the edge device transmits the selected patches to the server for an inference from the base model.  

    Another key metric, min-entropy, evaluates uncertainty in the most conservative manner~\cite{Konig2009operational}. 
    The min-entropy is defined as
    \begin{equation}
        g_m(\vect{x}) = -\log_2{\max_{y\in\mathcal{Y}}{p_\theta (y \vert \vect{x})}},
    \end{equation}
    which is directly associated with the confidence level of the initial inference. 
    If $g_m(\vect{x}) \ge \eta_m$, the edge device transmits the selected patches to the server for an inference from the base model.   

    Our experimental results in Section~\ref{sec:entropy_measure} show that the min-entropy serves as a better metric within our collaborative inference framework. 
    Entropy-aware image transmission utilizing the min-entropy improves communication efficiency for a given level of classification accuracy when compared to using the Shannon entropy.

    The entropy has been utilized in diverse applications, such as prioritizing unlabeled data inputs in active learning~\cite{Settles2012active} and optimizing wireless data acquisition for edge learning~\cite{Liu2021wireless}. 
    These works typically enhance training procedures using the entropy values calculated by complicated server models. 
    In contrast, our approach utilizes the min-entropy to assess the uncertainty of initial inferences made by a tiny model on the edge device instead of a complex server model. 
    This use of uncertainty metrics aims to minimize unnecessary image transmissions, thereby reducing communication overhead. 


    \section{Experimental Results}\label{sec:results}
    
    \subsection{Experiment Settings}

    Our experiments employ the ImageNet validation dataset and resize each image to a resolution of $224\times224$ pixels by center cropping. 
    An image is flattened to $N=196$ patches before the inference.

    We deploy DeiT-Tiny on the edge device and DeiT-Base on the server since resource-constrained edge devices have challenges in employing complicated models such as DeiT-Base. 
    The model complexity of these models is compared in Table~\ref{tab:model}. 
    DeiT-Tiny can be viable for deployment on edge devices such as NVIDIA Jetson Nano~\cite{Xu2023devit}, Raspberry Pi 4B~\cite{Dong2023packqvit}, and iPhone 12~\cite{Mehta2022mobilevit}. 
    In contrast, DeiT-Base is generally considered inappropriate due to its substantial memory consumption and prolonged inference latency. 
    For instance, the authors of~\cite{Xu2023devit} explicitly state that DeiT-Base is inadequate for deployment on NVIDIA Jetson Nano due to its substantial memory consumption and computational complexity. 
    Similarly, the authors of~\cite{Mehta2022mobilevit} emphasize memory consumption as a critical factor for edge device deployment, considering only lightweight models such as DeiT-Tiny and PiT~\cite{Heo2021rethinking} with approximately 3--6 million parameters.

    We evaluate the impact of varying the number of transmitted patches on communication cost and classification accuracy. 
    The communication cost in our collaborative inference system is quantified by the ratio of the number of transmitted patches to the total number of patches. 
    Consequently, if the edge device sends all image patches to the server, the communication cost is quantified as $1$.

    When the edge device transmits the selected patches, it is required to transmit the position information of these selected patches. 
    For instance, assigning one bit per patch as a marker of its selection status is a practical solution. 
    The overhead of these additional bits for position information is negligible in comparison to the size of the image itself, given that only one bit is appended for each image patch containing $6,144$ bits.

    \subsection{Communication Cost vs. Classification Accuracy}\label{sec:mainresult}

    In our collaborative inference framework, we assess the trade-off between communication cost and classification accuracy. 
    We utilize the attention-sum threshold selection for attention-aware patch selection and the min-entropy for entropy-aware image transmission. 
    
    \begin{figure}[!hbt] 
        \centering
        \includegraphics[width=0.4\textwidth]{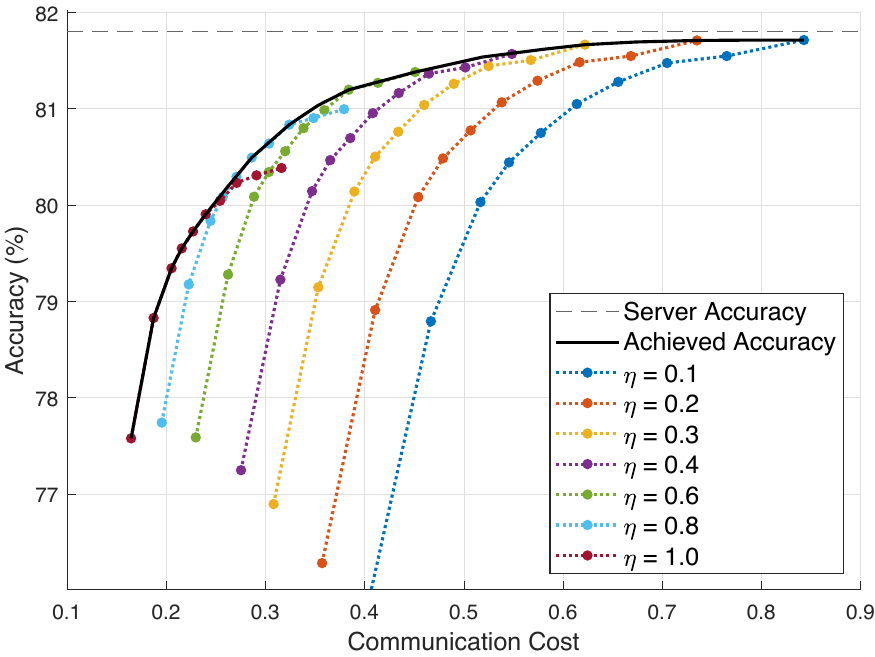}
        \caption{Trade-off between communication cost and classification accuracy, with the edge device employing DeiT-Tiny and the server employing DeiT-Base. The black line represents the achievable curve by the proposed collaborative inference. We utilize the attention-sum threshold selection method for attention-aware patch selection. For entropy-aware image transmission, the min-entropy serves as the entropy metric, using a threshold value $\eta = \eta_m$.}
        \label{fig:mainresult}
    \end{figure} 

    \begin{figure}[t] 
      \centering
      \includegraphics[width=0.4\textwidth]{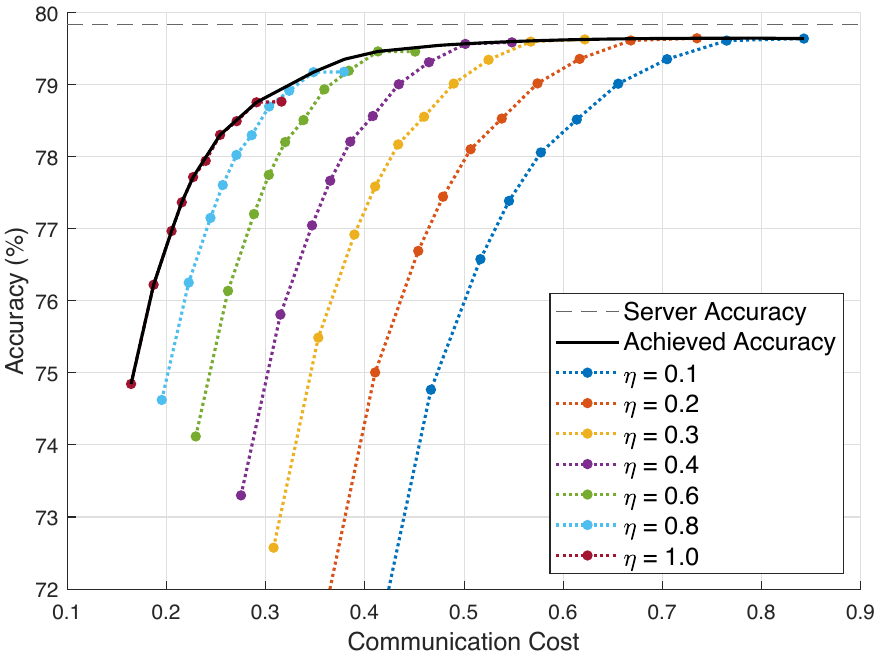}
      \caption{Trade-off between communication cost and classification accuracy, with the edge device employing DeiT-Tiny and the server employing DeiT-Small. The black line represents the achievable curve by the proposed collaborative inference. We utilize the attention-sum threshold selection method for attention-aware patch selection. For entropy-aware image transmission, the min-entropy serves as the entropy metric, using a threshold value $\eta = \eta_m$.}
      \label{fig:mainresult_deit_s}
      \vspace{-4mm}
   \end{figure}

    Fig.~\ref{fig:mainresult} shows the trade-off between communication cost and classification accuracy, employing DeiT-Tiny on the edge device and DeiT-Base on the server. 
    We achieve a \SI{68}{\%} reduction in communication cost while attaining a classification accuracy of \SI{80.84}{\%}, with only a minimal loss in accuracy compared to DeiT-Base's accuracy of \SI{81.8}{\%}.
    This communication cost reduction is achieved by the attention-sum threshold selection method with a threshold $\delta_\text{sum} = 0.97$ and the min-entropy with the threshold $\eta_m = 0.8$. 
    The black line indicates the optimized trade-off curve achieved by selecting the optimized threshold values of $\delta_\text{sum}$ and $\eta_m$. 
    Table~\ref{tab:sumth} and Table~\ref{tab:minent} detail the resulting communication costs depending on threshold values. 
    
    By controlling the threshold values of $\eta$ and $\delta$, our proposed collaborative inference framework can effectively manage communication resources. 
    In scenarios with reliable communication channels, lower values of $\eta$ and $\delta$ can be selected to increase data transmissions and maximize classification accuracy. 
    When the channel quality is worse and communication resources are limited, higher values of $\eta$ and $\delta$ can be set to balance classification accuracy with the available communication resource budget. 
    This strategy offers flexibility in adapting to varying communication channel conditions, thereby enhancing the efficiency of communication resource management. 
    
    \begin{table}[t]
        \centering
        \caption{Attention-sum Patch Selection Threshold Values and Expected Number of Transmitted Patches} 
        \renewcommand{\arraystretch}{1.2}
        \begin{tabular}{|c|c|}
            \hline
            $\delta_\text{sum}$ & Expected Number of Transmitted Patches\\ \hline \hline
             0.87 & 88.40 \\
             0.9 & 100.84  \\
             0.92 & 110.82 \\
             0.93 & 116.51 \\ 
             0.94 & 122.80  \\
             0.95 & 129.86  \\
             0.96 & 137.89  \\
             0.97 & 147.24  \\ 
             0.98 & 158.47  \\ 
             0.99 & 172.76  \\ \hline             
        \end{tabular}
        \label{tab:sumth}
    \end{table}
    
    \begin{table}[t]
        \centering
        \caption{Min-entropy Threshold Values and Expected Ratio of Transmitted Images}        
        \renewcommand{\arraystretch}{1.2}
        \begin{tabular}{|c|c|}
            \hline
            $\eta_m$ & Expected Ratio of Transmitted Images \\ \hline \hline
             1 & 0.3567 \\
             0.8 & 0.4290 \\
             0.6 & 0.5116 \\
             0.4 & 0.6246 \\ 
             0.3 & 0.7109 \\
             0.2 & 0.8445 \\
             0.1 & 0.9714 \\ \hline             
        \end{tabular}
        \vspace{-4mm}
        \label{tab:minent}
    \end{table}
    

   \begin{figure}[t] 
      \centering
      \includegraphics[width=0.4\textwidth]{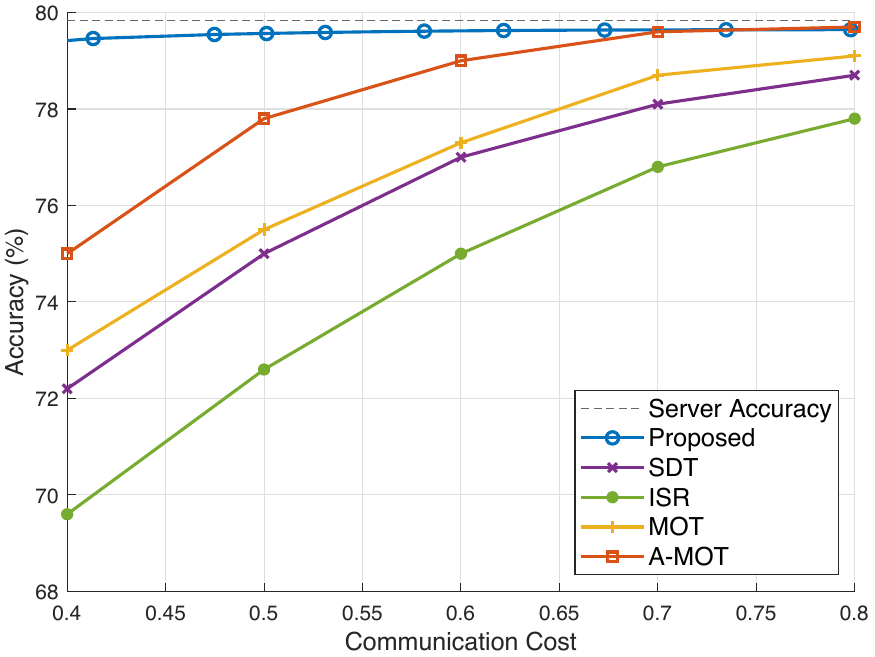}
      \caption{
      Comparison of previous methods and our collaborative inference framework, utilizing DeiT-Tiny on the edge device and DeiT-Small on the server. 
      The experiment is conducted on the ImageNet dataset.}
      \label{fig:mainresult_mot}
      \vspace{-4mm}
   \end{figure}

    Additionally, we investigate a case where the server employs DeiT-Small instead of DeiT-Base. 
    Fig.~\ref{fig:mainresult_deit_s} shows the trade-off between communication cost and classification accuracy. 
    In this case, we achieve a \SI{71}{\%} reduction in communication cost while attaining a classification accuracy of \SI{78.8}{\%}, with only a minimal loss in accuracy compared to DeiT-Small's accuracy of \SI{79.8}{\%}.
    It is observed that greater communication cost reductions can be realized when the accuracy gap between the edge device and the server is reduced.
    
   
    Fig.~\ref{fig:mainresult_mot} compares our collaborative inference framework with several existing methods, including server-driven transmission (SDT)~\cite{Du2020server}, image super-resolution (ISR)~\cite{Wang2021real}, masked autoencoder (MAE)-based offloading for transformer inference (MOT)~\cite{Liu2023efficient}, and adaptive MOT (A-MOT)~\cite{Liu2024adaptive}. 
    SDT transmits low-quality images initially and then offloads high-quality content from the target area based on the server's feedback~\cite{Du2020server}. 
    ISR transmits a low-quality image and then reconstructs a high-resolution image by super-resolution method~\cite{Wang2021real}. 
    The server model performs inference on this reconstructed high-resolution image. 
    In MOT and A-MOT, the edge device \emph{randomly} selects image patches for transmission to the server. 
    The server then reconstructs the entire image using the decoder of MAE~\cite{He2022masked} and performs classification on this reconstructed image.    
    The experimental results show that our proposed collaborative inference framework significantly improves performance by employing the tiny model on the edge device, which adeptly transmits essential image patches for classification. 

    \subsection{Attention-aware Patch Selection}\label{sec:patch_selection}

    In this subsection, we delve into identifying the effective attention metrics for determining patch importance and the optimal rules for patch selection. 
    First, we examine and compare the efficacy of mean attention scores versus attention rollout. 
    Next, we evaluate different patch selection methodologies, including top-$k$ selection, attention threshold selection, and attention-sum threshold selection. 
    We maintain a consistent environment with DeiT-Tiny on the edge device and DeiT-Base on the server. 
    For this analysis, we specifically focus on attention-aware patch selection, excluding considerations of entropy-aware image transmission.

    \begin{figure}[!t] 
        \centering
        \subfloat[]{\includegraphics[width=0.4\textwidth]{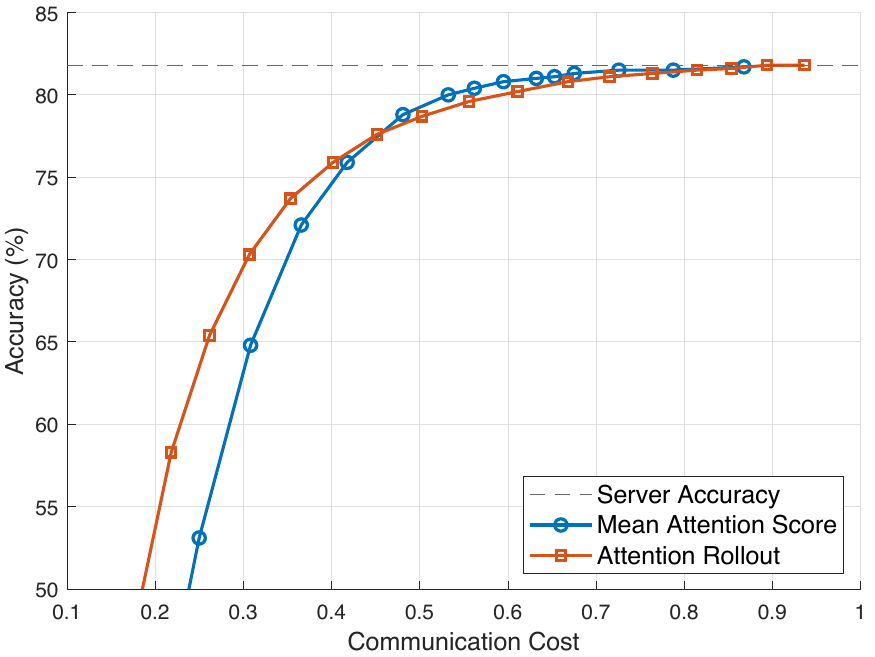}\label{fig:attention-measure-a}}
        \hfil
        \subfloat[]{\includegraphics[width=0.4\textwidth]{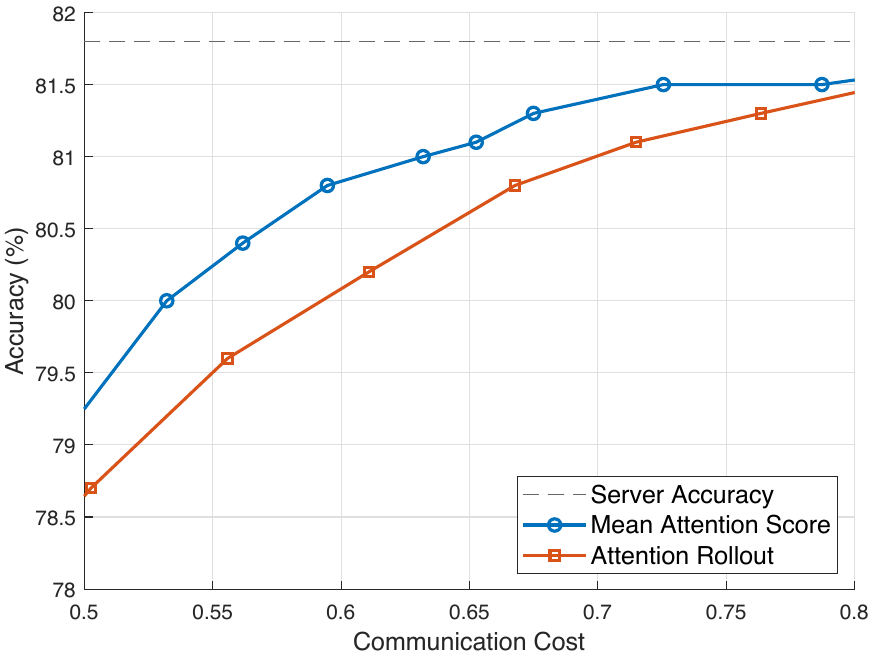}\label{fig:attention-measure-b}}
        \hfil
        \subfloat[]{\includegraphics[width=0.4\textwidth]{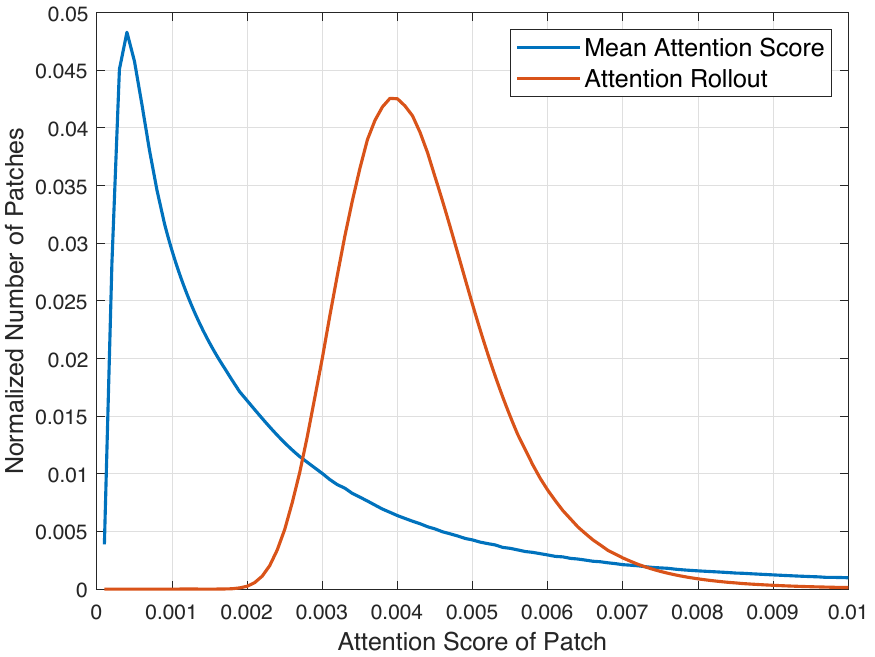}\label{fig:attention-measure-c}}
        \caption{Comparison of mean attention score and attention rollout, with the edge device employing DeiT-Tiny and the server employing DeiT-Base on the ImageNet dataset. The patch selection rule is the attention-sum threshold selection. (a) shows the overall trade-off between communication cost and classification accuracy, while (b) zooms in on the region of interest, focusing on areas near the server model's classification accuracy. (c) shows the normalized histograms of mean attention score and attention rollout, obtained by DeiT-Tiny. }
        \label{fig:attention-measure}
        \vspace{-4mm}
    \end{figure} 
    

    Fig.~\ref{fig:attention-measure}\subref{fig:attention-measure-a} reveals that both the mean attention score and attention rollout have distinct advantages. 
    In situations demanding substantial reductions in communication cost, attention rollout is better than the mean attention score. 
    However, as shown in Fig.~\ref{fig:attention-measure}\subref{fig:attention-measure-b}, the mean attention score is a better metric for attaining accuracy comparable to that of the server. 
   
    The theoretical advantage of the mean attention score over attention rollout is shown in Fig.~\ref{fig:attention-measure}\subref{fig:attention-measure-c}, which compares the normalized histograms of attention scores for both metrics. 
    Unlike attention rollout, which considers attention scores from multiple layers resulting in a more uniform distribution of scores, the mean attention scores concentrate on fewer patches with significantly higher relevance to the objects to be classified. 
    This concentration of attention scores is preferred because it implies that fewer but more relevant patches can be transmitted without compromising the classification accuracy, thus aligning with our objective to minimize communication overhead while maintaining classification accuracy.
 
    Hence, we adopt the mean attention score for our primary experiments. 

    \begin{figure}[!t]
        \centering
        \includegraphics[width=0.4\textwidth]{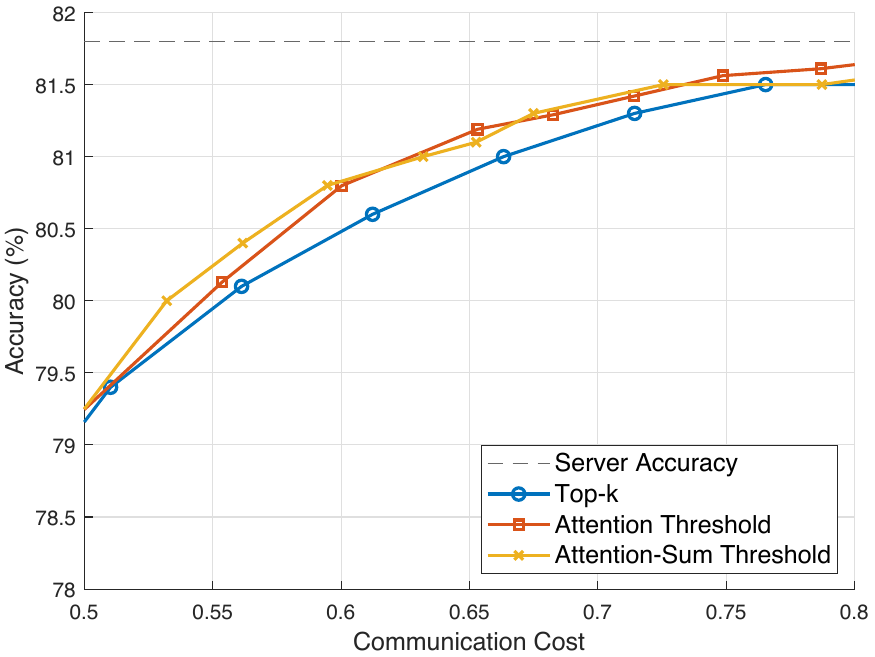}
        \caption{Comparison of the patch selection rules: Top-$k$ selection, attention threshold selection, and attention-sum threshold selection. The edge device and the server employ DeiT-Tiny and DeiT-Base, respectively. The patch importance is quantified by the mean attention score.}
        \label{fig:selection-rule}
    \end{figure}
   
    Fig.~\ref{fig:selection-rule} shows that both the attention threshold selection and attention-sum threshold selection outperform the top-$k$ selection.
    These methods offer the capability to adjust the number of selected patches for transmission, facilitating the maintaining classification accuracy while minimizing communication costs.    
  
    The variability in informational content across patches, influenced by factors such as object size, type, and the presence of background, directly impacts their importance for accurate classification.
    Theoretically, the attention scores reflect each patch’s contribution to the model prediction. 
    By setting thresholds based on cumulative attention scores (attention-sum threshold) or individual patch attention scores (attention threshold), we can dynamically control the number of transmitted patches based on their estimated relevance.

    \subsection{Entropy-aware Image Transmission}\label{sec:entropy_measure}

    \begin{figure}[t] 
        \centering
        \includegraphics[width=0.4\textwidth]{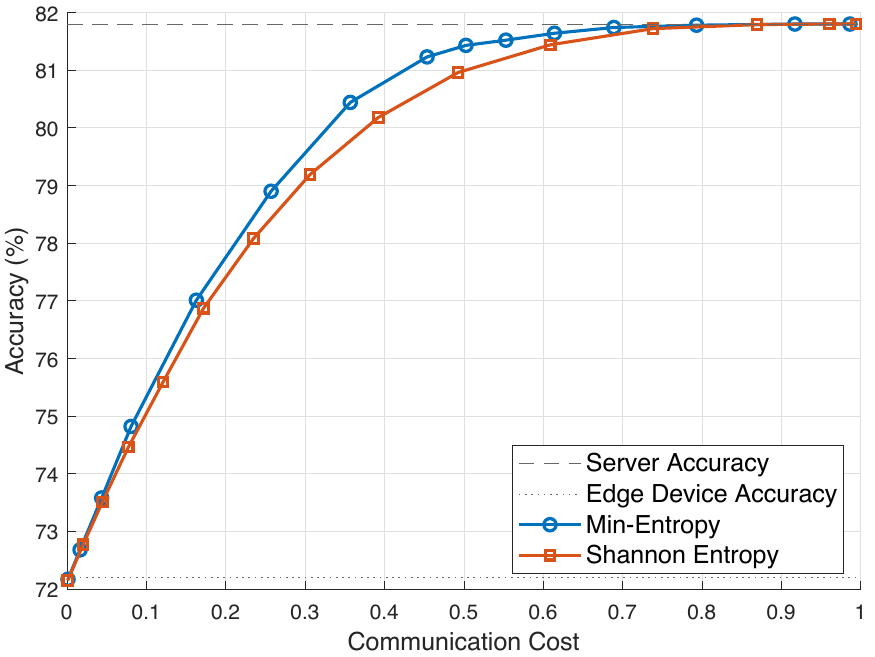}
        \caption{Comparison of min-entropy and Shannon entropy with the edge device employing DeiT-Tiny and the server employing DeiT-Base. }
        \label{fig:uncer-acc}
        \vspace{-4mm}
    \end{figure} 

    To determine the most effective entropy measure, we compare the min-entropy and the Shannon entropy within our collaborative inference framework. 
    For this analysis, we specifically focus on entropy-aware image transmission, deliberately setting aside the aspect of attention-aware patch selection.  
    The entropy values are derived from the softmax output of the MLP classification head in the DeiT-Tiny model used on the edge device. 
    As shown in Fig.~\ref{fig:uncer-acc}, the min-entropy is a better metric for our collaborative inference, demonstrating a more effective distinction between correctly and incorrectly inferred images compared to the Shannon entropy. 
    Consequently, min-entropy is chosen as the preferred metric for entropy-aware image transmission.
    
    \subsection{Comparison of Models as Semantic Encoder }\label{sec:semantic_enc} 
   
\begin{figure}[t] 
      \centering
      \includegraphics[width=0.4\textwidth]{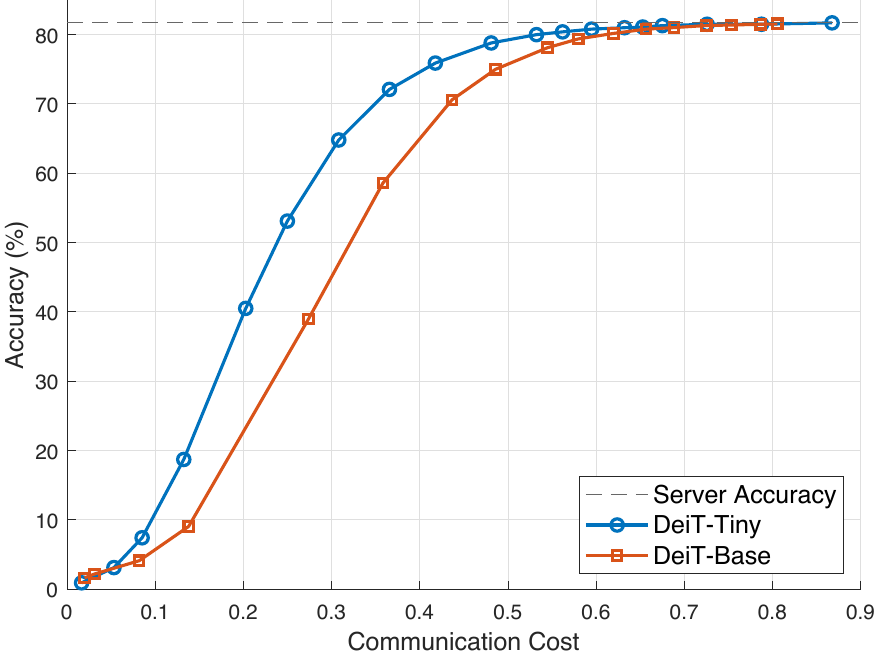}
      \caption{Comparison of DeiT-Tiny and DeiT-Base as semantic encoders to select important patches. The classification accuracy is obtained by DeiT-Base.}
      \label{fig:model-size}
   \end{figure}

    As discussed in Section~\ref{sec:quantifying}, DeiT-Tiny can act more effectively as a semantic encoder than DeiT-Base in spite of its inferior classification accuracy.
    In our experiments, we concentrate solely on attention-aware patch selection to evaluate and compare DeiT-Tiny and DeiT-Base. 
    Both models select only important patches based on the mean attention scores, disregarding initial inference outcomes. 
    The classification accuracy is obtained by DeiT-Base, which processes only these selected image patches. 
    Fig.~\ref{fig:model-size} shows that DeiT-Tiny more effectively identifies the essential image patches for classification better than DeiT-Base.

    In particular, Fig.~\ref{fig:attention_model} and Fig.~\ref{fig:attention_map_model} in Section~\ref{sec:quantifying} show that the attention scores obtained by DeiT-Tiny concentrate more on relevant objects than those from DeiT-Base. 
    Recent work ~\cite{Darcet2024vision} supports this observation, showing that large ViT models often allocate high attention scores to less informative background areas. 
    It is because the large ViT models adeptly identify patches containing minimal information, such as background areas, repurposing the corresponding patches to assimilate global image information while neglecting spatial information.
    While these high attention score patches may enhance the classification accuracy, they diminish the effectiveness of base models as semantic encoders. 

    \subsection{Inference Latency Analysis}\label{sec:latency}

    In this subsection, we evaluate the end-to-end inference latency of both our collaborative inference and the conventional server-based inference.

    The end-to-end inference latency consists of client inference latency, server inference latency, and communication latency. 
    Client inference latency, set as 10.99 milliseconds (ms), reflects the inference latency of DeiT-Tiny on the iPhone 12 neural engine~\cite{Mehta2022mobilevit}. 
    Server inference latency is set at 8.32 ms, corresponding to the inference latency of DeiT-Base on an NVIDIA RTX 3090 GPU~\cite{Xu2023devit}. 
    This latency can be eliminated if the image is not transmitted to the server via the strategy of entropy-aware image transmission. 
    Additionally, server inference latency can be reduced in proportion to the decreased computational complexity (FLOPs) if only selected patches are transmitted via the strategy of attention-aware patch selection. 
    The FLOPs of DeiT-Base are calculated by \eqref{eq:complexity}. 
    Communication latency is calculated using the transmitted data size divided by the upload data rate. 
    We estimate the data size of a typical image to be 147 KB, based on an image cropped to $224\times224\times3$ bytes. 
    The upload data rates of 1 Mbps, 8 Mbps, and 20 Mbps are considered as in~\cite{Zhang2023effect}.

    \begin{figure}[t] 
        \centering
        \includegraphics[width=0.4\textwidth]{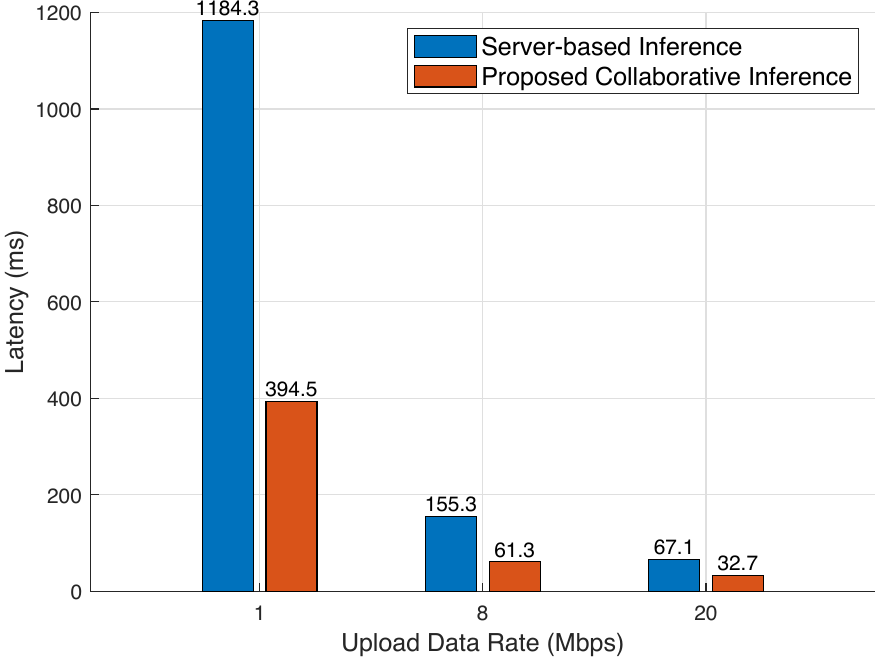}
        \caption{Comparison of inference latency for different upload data rates, 1 Mbps, 8 Mbps, and 20 Mbps.}
        \label{fig:system_latency}
    \end{figure}

    \begin{figure}[t] 
        \centering
        \includegraphics[width=0.4\textwidth]{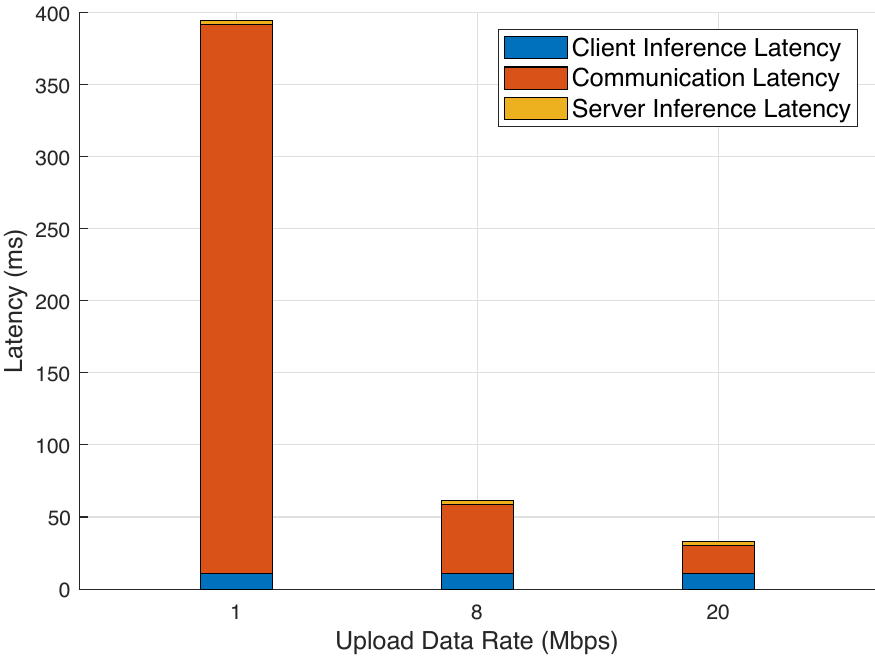}
        \caption{Breakdown of end-to-end inference latency of the proposed collaborative inference.}
        \label{fig:type_latency}
    \end{figure}

    Fig.~\ref{fig:system_latency} compares the inference latency of the server-based inference and our proposed inference for each upload data rate. Across these rates, our proposed inference framework consistently exhibits better inference latency. 
    We set the thresholds $\delta_\text{sum}$ and $\eta_m$ at values that result in a minimal accuracy loss of \SI{1}{\%}, as described in Section~\ref{sec:mainresult}.
    Fig.~\ref{fig:type_latency} shows the proportions of client inference latency, server inference latency, and communication latency. Notably, communication latency occupies a significant portion at lower upload data rates.


    \section{Conclusion}\label{sec:conclusion}
    
    We presented a novel attention-aware collaborative inference framework using pre-trained ViT models. 
    The edge device employs a lightweight ViT model as a semantic encoder, selecting pivotal patches that focus on objects crucial for classification. 
    Our results confirm that this framework not only significantly lowers communication costs but also preserves accuracy comparable to that of server models. 
    Furthermore, it offers the added benefit of diminishing the server model's computational complexity. 
    Extending this approach to encompass a broader range of tasks beyond classification provides an intriguing direction for future research.
    
    \appendices
	
    \bibliographystyle{IEEEtran}
    \bibliography{abrv,mybib}

\end{document}